\documentclass[11pt]{article}
\pdfoutput=1
\usepackage{latexsym}
\usepackage{amssymb}
\usepackage{amsmath}
\usepackage{graphicx}

\newcommand{\lsim}{\lesssim}

\begin{document}
\pagestyle{empty}

\vspace*{1cm}

\begin{center}

{\bf \LARGE A Holographic Realization of Ferromagnets} 
\\

\vspace*{2cm}
{\large 
Naoto Yokoi$^{1}$\footnote{yokoi@imr.tohoku.ac.jp}, Masafumi Ishihara$^{2}$\footnote{masafumi@wpi-aimr.tohoku.ac.jp}, Koji Sato$^{2}$\footnote{koji.sato@wpi-aimr.tohoku.ac.jp}, and Eiji Saitoh$^{1, 2, 3, 4}$\footnote{eizi@imr.tohoku.ac.jp}} \\
\vspace*{1.5cm}

{\it $^{1}$Institute for Materials Research, Tohoku University, Sendai 980-8577, Japan \\
$^{2}$WPI-Advanced Institute for Materials Research, Tohoku University, \\
Sendai 980-8577, Japan \\
$^{3}$ERATO, Spin Quantum Rectification Project, Japan Science and Technology Agency, Sendai 980-8577, Japan \\
$^{4}$Advanced Science Research Center, Japan Atomic Energy Agency, Tokai 319-1195, Japan}
\end{center}

\vspace*{3.5cm}

\begin{abstract}
{\normalsize
A holographic realization for ferromagnetic systems has been constructed.
Owing to the holographic dictionary proposed on the basis of this realization, we obtained
relevant thermodynamic quantities such as magnetization, magnetic
susceptibility, and free energy. This holographic model reproduces the
behavior of the mean field theory
near the critical temperature. At low temperatures, the results
automatically incorporate the contributions from spin wave excitations and
conduction electrons.}
\end{abstract} 

\newpage
\baselineskip=18pt
\setcounter{page}{2}
\pagestyle{plain}
\baselineskip=18pt
\pagestyle{plain}

\setcounter{footnote}{0}

\section{Introduction}
Spintronics is the field which aims at development of next generation devices by controlling both spin and charge degrees of freedom in solid state systems 
\cite{Zutic, Tserkovnyak, Brataas, Jungwirth, Bauer}, where many of these achievements 
were guided by fundamental principles such as symmetry and reciprocity. Recently, the holographic duality \cite{Mal, GKP, Witten}, which relates a $d$-dimensional quantum field theory to a classical gravitational system in ($d$+1) dimensions, has been applied to the field of condensed matter physics. A new guideline in spintronics could also emerge through such a duality.\footnote{A few applications of the holographic duality to spintronics have been discussed in \cite{Bigazzi, Hashimoto}.} Particularly, manipulation of spins is achieved by various non-equilibrium dynamics of magnetizations in ferromagnetic systems, and a number of phenomenological approaches are taken \cite{MaekawaBook}. The holographic duality can map thermal and dissipative phenomena in quantum field theory to corresponding phenomena in gravitational dynamics, and it can provide a novel approach to aforementioned non-equilibrium phenomena. (For a review, see Ref. \cite{Hubeny}.) 

In this Letter, towards applications of the holographic duality to magnetization dynamics, as a first step, we develop a holographic dictionary between an 
appropriate gravitational theory and a ferromagnetic system, where we can calculate relevant physical quantities such as magnetization and magnetic susceptibility at finite temperatures and chemical potentials. The holographic duality so far had been applied to two-dimensional magnetic systems in Refs.  \cite{Iqbal, Cai1, Cai2015}, where the behaviors near the critical temperature were discussed. Our setup deals with three-dimensional magnetic systems and describes their behavior not only near the critical temperature but also in low temperatures where the technology of spintronics is actively developed. Our holographic model in principle can provide a means to analyze phenomena involving not only magnetization but also spin transport, and thus it can introduce new perspectives in the field of spintronics.

\section{A Holographic Realization of Ferromagnets}
In order to construct a holographic dictionary, we propose a holographic dual model of three-dimensional ferromagnets described by the five-dimensional 
gravitational action:
 \begin{eqnarray}
S =\int\!\sqrt{- g}\, d^5 x  \left[ 
\frac{1}{2 \kappa^2}\left( R - 2 \Lambda \right) - \frac{1}{4 e^{2}} G_{M N} G^{M N} 
- \frac{1}{4 g^{2}} F_{M N}^{a} F^{a\, M N} - \frac{1}{2} \left(D_{M} \phi^{a}\right)^2 - 
V(|\phi|) \right]\,, \label{eq: Holo action}
\end{eqnarray}
where $R$ is the scalar curvature, and $\Lambda$ is the five-dimensional 
negative cosmological constant.
$\phi^{a}$ is a triplet scalar field of the $SU(2)$ gauge symmetry. We have introduced a $SU(2)$ invariant potential $V(|\phi|) = \lambda \left(|\phi|^2-m^2/\lambda\right)^2\!/4$, which is a quartic polynomial 
for simplicity.\footnote{$|\phi|$ is a norm for scalars given by $|\phi|^2 = \sum_{a=1}^{3}(\phi^{a})^{2}$.}  Here, $m$ and $\lambda$ are the parameters describing the mass and coupling strength, respectively. 
$a$ is the spin index ($a = 1 \sim 3$), and the five-dimensional spacetime index is labeled by $M$ and $N$  ($M, N = 0 \sim 4$). $G_{M N} = \partial_{M} B_{N} - \partial_{N} B_{M}$ and $F_{M N}^{a} = \partial_{M} A_{N}^{a} - \partial_{N} A_{M}^{a} + \epsilon^{a b c}A_{M}^{b}A_{N}^{c}$ give the field strength, and the covariant derivative is defined as $D_{M} \phi^{a} = \partial_{M} \phi^{a} + \epsilon^{a b c} A_{M}^{b} \phi^{c}$ with a totally anti-symmetric tensor with $\epsilon^{1 2 3} = 1$. Here, $B_{M}$ is a $U(1)$ gauge field and $A_{M}^{a}$ is a $SU(2)$ gauge field.  

In the context of holographic duality, the fields in the dual gravitational system correspond to the operators in the ferromagnetic system. The field-operator correspondence can be found based on the symmetry argument \cite{GKP,Witten}, and we can make a dictionary between the ferromagnets and the dual gravitational model. From the translational symmetry in the continuous limit, the conserved currents in ferromagnets are the stress tensor $T_{\mu\nu}~ (\mu, \nu = 0 \sim 3)$ which couples to the metric in the dual model (\ref{eq: Holo action}). Similarly, for the $SU(2)$ spin rotational symmetry, the conserved currents are spin currents $J_{\mu}^{a}$ which are identified with the $SU(2)$ gauge fields $A_M^a$ in the dual model. Magnetization $M^{a}$, which is a vector quantity under the $SU(2)$ rotation, is identified with the triplet scalar fields $\phi^a$ in the dual model. In addition, the $U(1)$ gauge field $B_{M}$ corresponds to the electric current $J_{\mu}$. Note that the triplet scalars, dual to the magnetization vector, are neutral under this $U(1)$ gauge symmetry. Our holographic dictionary can be summarized in the table below.
\begin{center}
\begin{tabular}{ccc}
Ferromagnet & & Dual gravity \\ \hline
Magnetization~ $M^{a}$ & $\Longleftrightarrow$ & Scalar field~ $\phi^{a}$\\
Spin current~ $J_{\mu}^{a}$ & $\Longleftrightarrow$ & $SU(2)$ gauge field~ $A_{M}^{a}$\\
Charge current~ $J_\mu$ & $\Longleftrightarrow$ & $U(1)$ gauge field~ $B_M$\\
Stress tensor~ $T_{\mu\nu}$ & $\Longleftrightarrow$ & Metric~ $g_{M N}$  
\end{tabular}
\end{center}
The GKPW relation \cite{GKP, Witten} in the holographic duality and the above dictionary give a prescription to calculate the correlation functions of the operators in ferromagnets using our dual gravitational model (\ref{eq: Holo action}).

\section{Charged Black Holes and Finite Temperatures and Chemical Potentials}
A typical feature of ferromagnetism is the existence of the phase transition between a paramagnetic and ferromagnetic phases at the Curie temperature. Some holographic models for systems exhibiting phase transition have been proposed, such as holographic superconductors \cite{HSC, HSC2}. Here, the notion of temperature is introduced by black holes as the background of the dual gravitational system \cite{Witten}. Our analysis of the ferromagnetic phase transition follows the similar setting. 

At first, we begin with a paramagnetic phase, where the expectation value of the magnetization vanishes, and thus $\phi^{a} = 0$ in the dual system.    
Then, one can analytically obtain the charged black hole solution \cite{su2}, which is invariant under translations and rotations in three-dimensional space $(x, y, z)$ (up to gauge rotation). The metric is given by
\begin{equation}
ds^2=\frac{r^2}{\ell^2} \left(-fd\tilde{t}^2 + dx^2 + dy^2 + dz^2 \right) 
+ \frac{\ell^2}{f} \frac{dr^2}{r^2}\,, \label{eq: BH metric}
\end{equation}
where
\begin{equation}
f=f(r)=1 + Q^{2} \left(\frac{r_{H}}{r}\right)^{6} - (1 + Q^2) \left(\frac{r_{H}}{r}\right)^{4}\,.\label{eq: horizon}
\end{equation}
Here, the parameter $Q$ is given by
\begin{equation}
Q^2 = \frac{2 \kappa^2}{3} \left(\frac{\mu^2}{e^2} + \frac{\mu_{s}^2}{g^2} \right)\,, \label{eq: chemical potential}
\end{equation}    
and the time component of the gauge fields\footnote{Here, we have imposed the regularity condition at the horizon on the solutions of gauge fields.} are given by $B_{0}=\mu\left(r_{H}/\ell \right)\left(1-r_{H}^2/r^2\right)$ and $A_{0}^{3} = \mu_{s}\left(r_{H}/\ell \right)\left(1 - r_{H}^2/r^2\right)$.  
The black hole solution (\ref{eq: BH metric}) has the  horizon at $r = r_{H}$, and 
the cosmological constant $\Lambda = - 6/\ell^2$ leads to an asymptotically AdS spacetime. 
The temperature of this charged black hole is given by $T = (2-Q^2)/2 \pi$, 
where we have set $r_{H} = \ell = 1$. 

Through the holographic duality, the charged black hole (\ref{eq: BH metric}) plays the role of a thermal bath with temperature $T$ in the dual ferromagnetic system, and the gauge field solutions give the electro-chemical potential $\mu$ and the spin accumulation $\mu_{s}$ (difference between the chemical potentials of up and down spins) \cite{MaekawaBook} in Eq. (\ref{eq: chemical potential}). With the charged black hole as the background in our dual gravitational theory, we can analyze the magnetization dynamics at finite temperatures and chemical potentials.

\section{Magnetization, Magnetic Susceptibility, and Specific Heat}
\begin{figure}[htb]
\begin{center} 
\includegraphics[width=80mm]{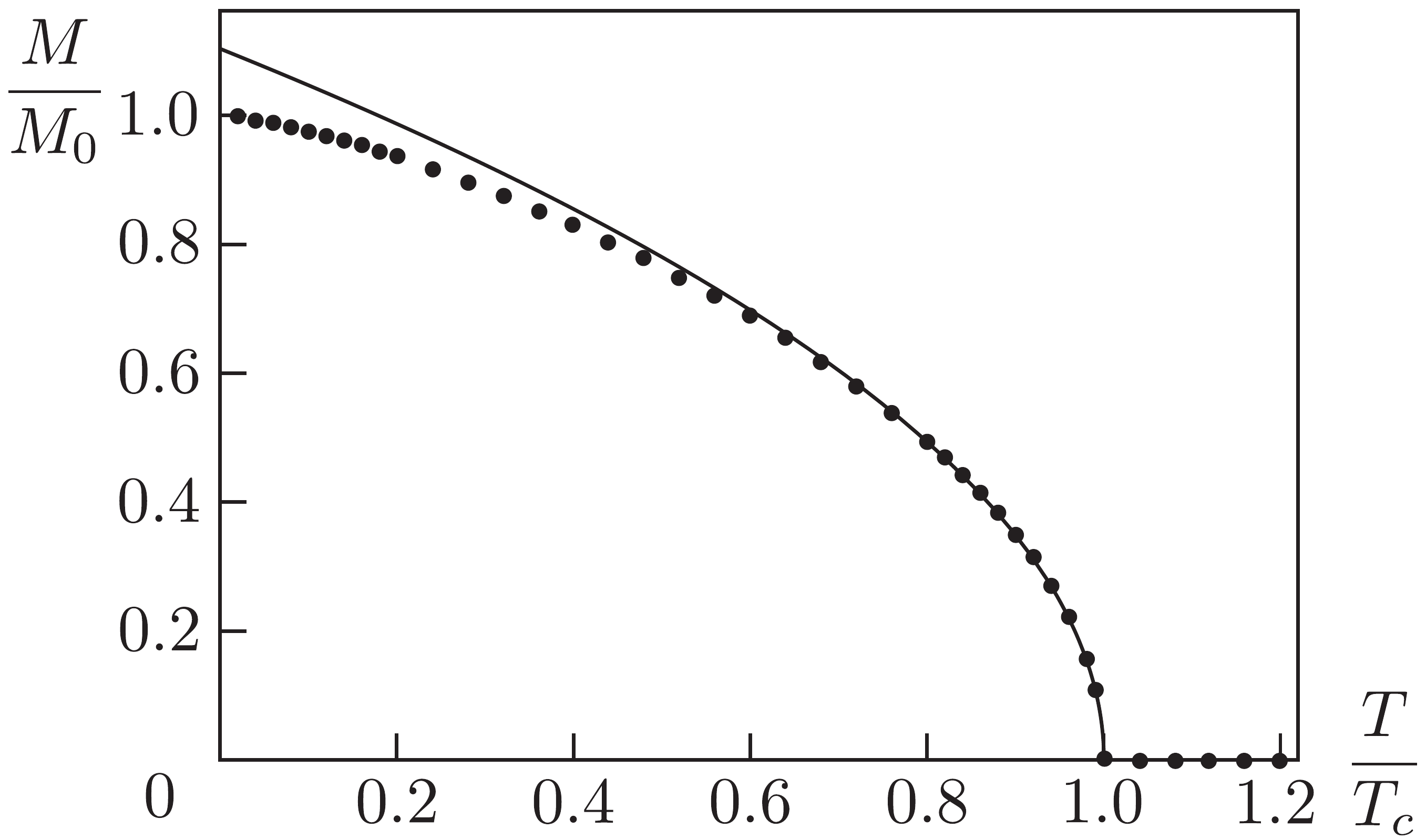}
\caption{Temperature dependence of normalized magnetization for $m^2 = 3.89$ (dots) and the fitting function (\ref{eq: fitfn magtc}) near $T_{c}$ with $\tilde{M}_{0} = 0.0603$ (solid line). Note that $M_0$ is  the magnetization at zero temperature in Eq. (\ref{eq: fitfn magzero}).}
\label{fig1}
\end{center}
\end{figure}
Using the holographic dictionary,   
finding the magnetic observables is reduced to solving the equations of motion of scalars $\phi^{a}$ in the charged black hole background given by Eq.  (\ref{eq: BH metric}). For a static and homogeneous magnetization, we can assume $\phi^{a}$ is non-vanishing only for $a=3$ and dependent only on $r$ without loss of generality. With these assumptions, the equation of motion of 
$\Phi \equiv \phi^{3}(r) /\sqrt{\lambda}$ is given by the following ordinary differential equation:        
\begin{equation}
r^2 f\, \Phi'' + \left(5\, r f + r^2 f' \right) \Phi' +m^2\Phi-\Phi^3  = 0\,. \label{eq: phi eom}
\end{equation}
This equation governs the static properties of the holographic dual ferromagnets.
Note that Eq. (\ref{eq: phi eom}) is completely separated from the equations of motion for other fields, in particular, gauge fields. 

According to the holographic recipe \cite{Balasubramanian, Klebanov}, the boundary behavior of a solution to Eq. (\ref{eq: phi eom}) gives an expectation value of the magnetization of the dual ferromagnetic system. 
Since the charged black hole background asymptotically approaches to the AdS spacetime near $r \sim \infty$, the solution has an asymptotic expansion in the form of
\begin{equation}
\Phi(r) = a_{0} \left(\frac{1}{r}\right)^{\Delta_{-}} +\, b_{0} \left(\frac{1}{r}\right)^{\Delta_{+}} + \cdots\,, \label{eq: asymptotic form} 
\end{equation}
where the exponents are given by $\Delta_{\pm} = 2 \pm \sqrt{4 - m^2}$. In this expansion, the coefficient $a_{0}$ corresponds to an external magnetic field, and $b_{0}$ corresponds to an expectation value 
of the magnetization.

In order to investigate the ferromagnetic phase transition with spontaneous symmetry breaking, 
we set the external magnetic field to zero ($a_{0} = 0$).  
Using a similar technique to the holographic superconductor \cite{HSC, HSC2}, 
we seek the numerical solution to Eq. (\ref{eq: phi eom}) with the asymptotic behavior 
(\ref{eq: asymptotic form}) under the condition $a_{0} = 0$. For a given temperature $T$, 
we find the coefficient $b_{0}$ in Eq. (\ref{eq: asymptotic form}) and identify it 
with a spontaneous magnetization $M(T)$. We have observed the phase transition 
and obtained the Curie temperature  $T_c$ of the dual ferromagnetic system 
with the parameters $3.5 \lsim m^2 \lsim 4$. 
For example, the numerical results of $M(T)$ are shown in 
Fig.\,\ref{fig1} for $m^2 = 35/9 \simeq 3.89$, where  $T_c$ is given by $T_{c} \simeq 8.76 \times 10^{-4}$. For the temperature above $T_{c}$, the solutions become identically $\Phi = 0$, corresponding to vanishing magnetizations.

By the same technique, we can also obtain the solutions with slightly non-vanishing values of $a_{0}$ in Eq. (\ref{eq: asymptotic form}), which lead to the values of magnetization under small external magnetic field $H$. From such solutions, the magnetic susceptibility can be evaluated at finite temperatures by $\chi_{H} (T) = \left. \frac{\partial M}{\partial H} \right|_{H=0} = \left. \frac{\partial b_{0}}{\partial a_{0}} \right|_{a_{0}=0}$.

Another static quantity which can be calculated from the holographic technique is the specific heat $c(T)$. The free energy of the ferromagnetic system $\beta F(T) \left(\equiv \frac{F(T)}{k_BT}\right)$ is identified with the value of the action pertaining to the scalar field in (\ref{eq: Holo action}), which is evaluated with the numerical solution of $\Phi$ in Eq.  (\ref{eq: phi eom}). Then, the specific heat is given by $c = -T\,  \partial^{2} F/\partial T^{2}$.

\section{Properties of Ferromagnetic Phase Transition}
First, we find the temperature dependence of magnetization near $T_c$ has a simple form given by
\begin{equation}
M(T)\, \simeq\, \tilde{M}_{0} \left( 1 - t \right)^{1/2}\,, \label{eq: fitfn magtc}
\end{equation}
where we defined the dimensionless temperature as $ t \equiv T/T_c$. 
The data and fitting plot near  $T_c$ are shown in Fig.\,\ref{fig1} for $m^{2} = 3.89$. This form of temperature dependence matches with the behavior of magnetization obtained from Ginzburg-Landau theory. Thus, our holographic dual model reproduces the well-known behavior of magnetization near  $T_c$. 

\begin{figure}[htb]
  \begin{center}
    \includegraphics[width=80mm]{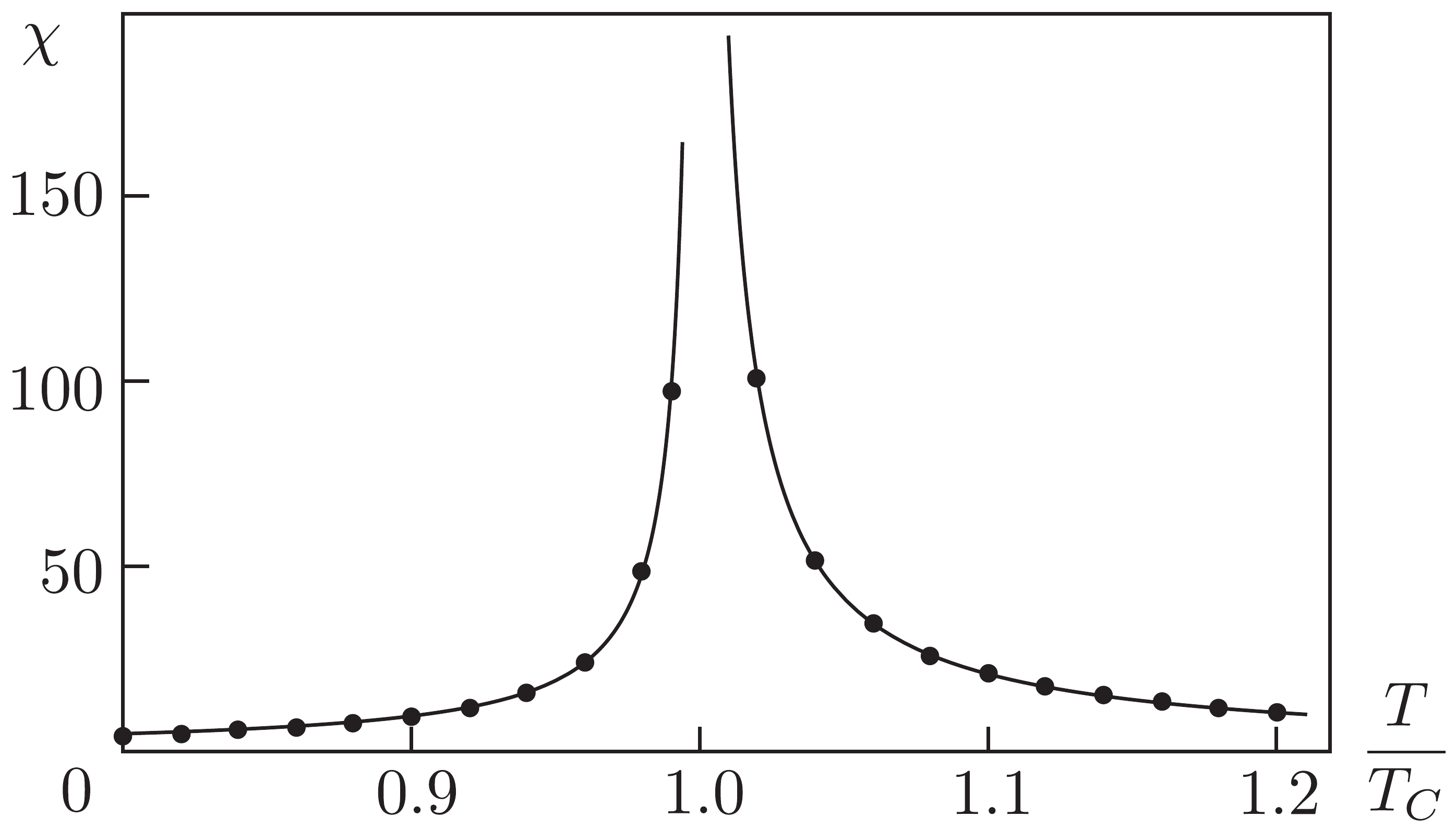}
  \end{center}
\caption{Magnetic susceptibility near  $T_c$ for $m^2=3.89$ and the functions (\ref{eq: Curie Weiss}) with $C_{\rm b} = 0.950$ and $C_{\rm a} = 2.11$.}
\label{fig2}
\end{figure}   
Next, the numerical results of temperature dependence of the susceptibility near  $T_c$ is summarized in Fig.\,\ref{fig2} for $m^2=3.89$.  As expected, the susceptibility $\chi_{H}(T)$ diverges at  $T_c$, and the behavior around $T_{c}$ is well fitted with the functions,
\begin{equation}
\chi_{H} (T)\, \simeq
\left\{\begin{array}{ll}
C_{\rm b}(1 - t)^{-1}  & \quad\left(t <  1 \right)\\
C_{\rm a}(t - 1)^{-1}  & \quad\left(t > 1 \right)
\end{array}\right.\,, \label{eq: Curie Weiss}
\end{equation}
where $C_{\rm b}$ and $C_{\rm a}$ are constants in front of the temperature dependent part of the susceptibility below and above $T_c$, respectively. This behavior is consistent with the Curie-Weiss law for the magnetic susceptibility near the phase transition point. Interestingly, we find that the ratio of the constants  $R = C_{\rm a}/C_{\rm b}$ takes its values in the range of $2 \lsim R \lsim 3$ for $3.5 < m^2 < 4$ (e.g., $R = 2.22$ for $m^2=3.89$). These values are close to the ratio $R = 2$ obtained from Ginzburg-Landau theory, and the small deviations may indicate our holographic dual approach is slightly beyond the mean field approximation.  

\begin{figure}
\begin{center} 
\includegraphics[width=80mm]{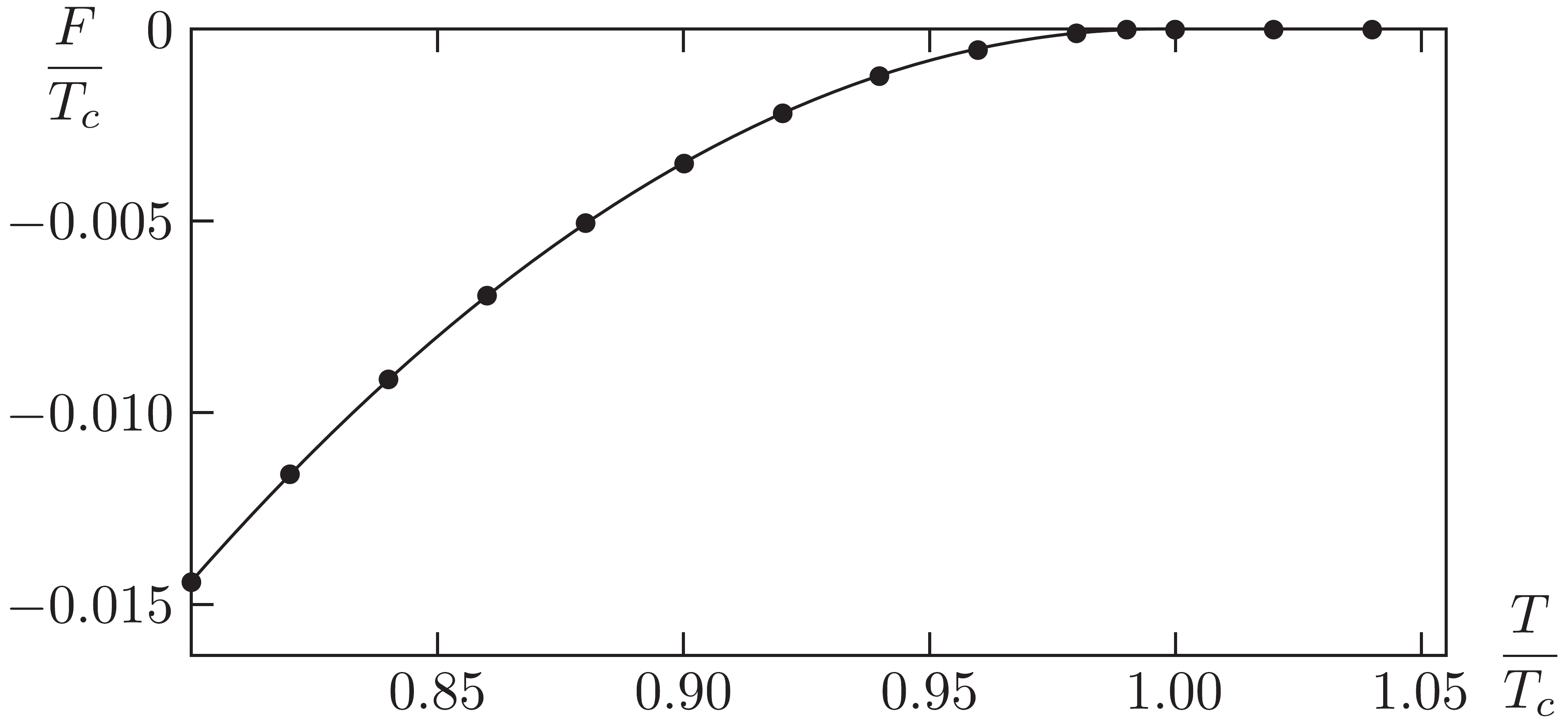}
\caption{Temperature dependence of free energy near $T_{c}$ for $m^2 = 3.89$ and $\lambda=1$ and the fitting function (\ref{eq: capacitytc}) with $C_{0}=3.23 \times 10^{-4}$.}
\label{fig3}
\end{center}
\end{figure}
Finally, the behavior of the free energy $F(T)$ near  $T_c$ turns out to be the following form
\begin{equation}
F(T)\,  \simeq\, -  C_{0} \left(1 - t \right)^2 \quad \quad \left( t \lsim 1 \right)\,, 
\label{eq: capacitytc}
\end{equation}
where $C_{0}$ is a positive constant. The numerical results for $m^2=3.89$ are shown in Fig.\,\ref{fig3}.

In summary, all the critical exponents found in our numerical results  $c(T) \simeq (1 - t)^{0}$, 
$M(T) \simeq (1 - t)^{1/2}$, and $\chi_{H}(T) \simeq |1 - t|^{- 1}$ are perfectly 
consistent with Ginzburg-Landau theory. 
We have also numerically confirmed the behavior $M \simeq H^{1/3}$ with external magnetic fields $H$ 
at $T=T_{c}$.  These results confirm the validity of the holographic dual model near the phase transition point.

\section{Magnetic Properties at Low Temperatures}  
\begin{figure}
\begin{center} 
\includegraphics[width=80mm]{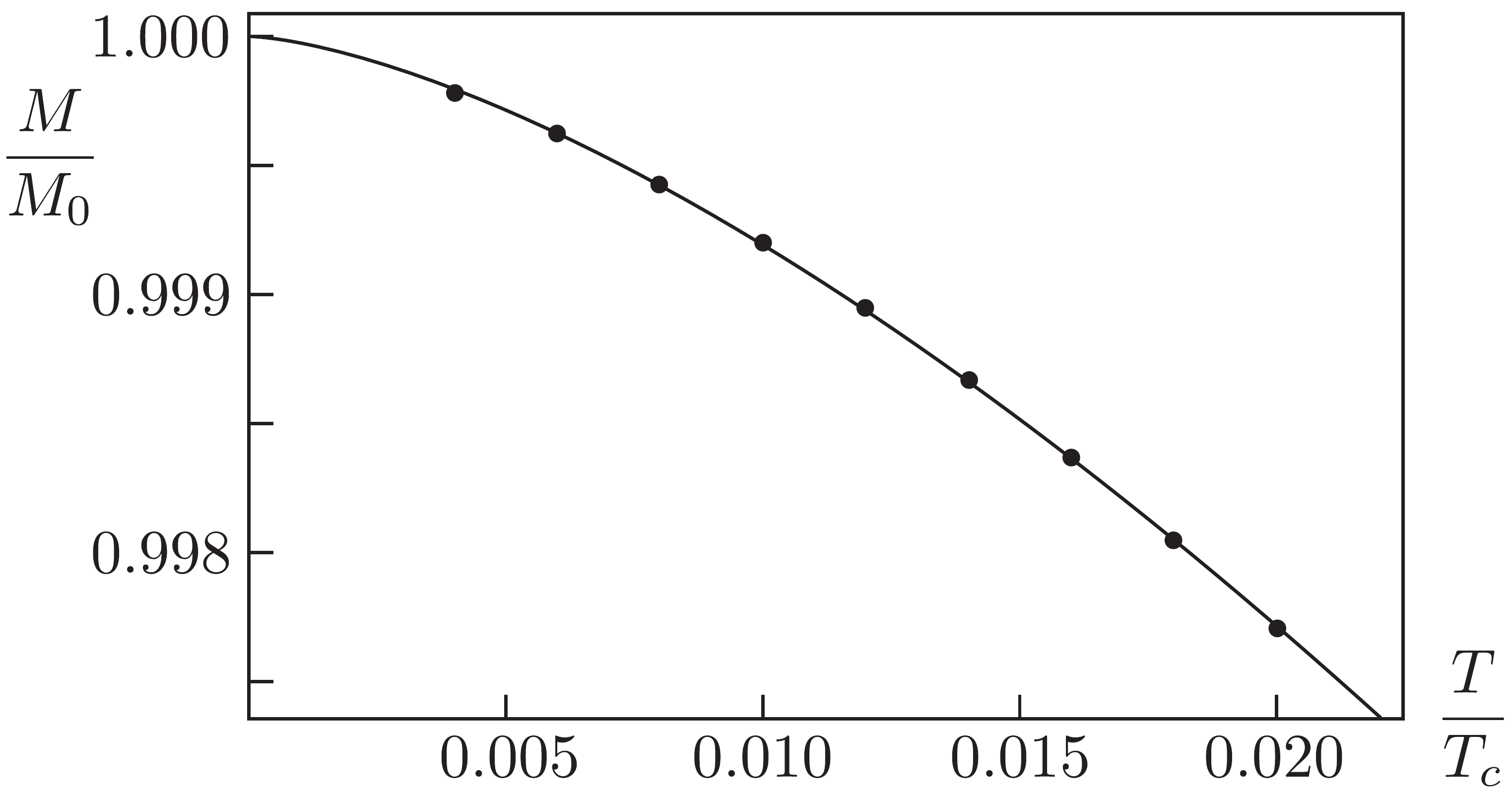}
\caption{Temperature dependence of magnetization in the low temperature for $m^2 = 3.89$ and the fitting function (\ref{eq: fitfn magzero}) with $M_{0} = 0.0546$ and $D_{0} = 0.808$.}
\label{fig4}
\end{center}
\end{figure}   
 \begin{figure*}[htb]
\begin{minipage}{0.9\hsize}
  \begin{center}
    \includegraphics[width=150mm]{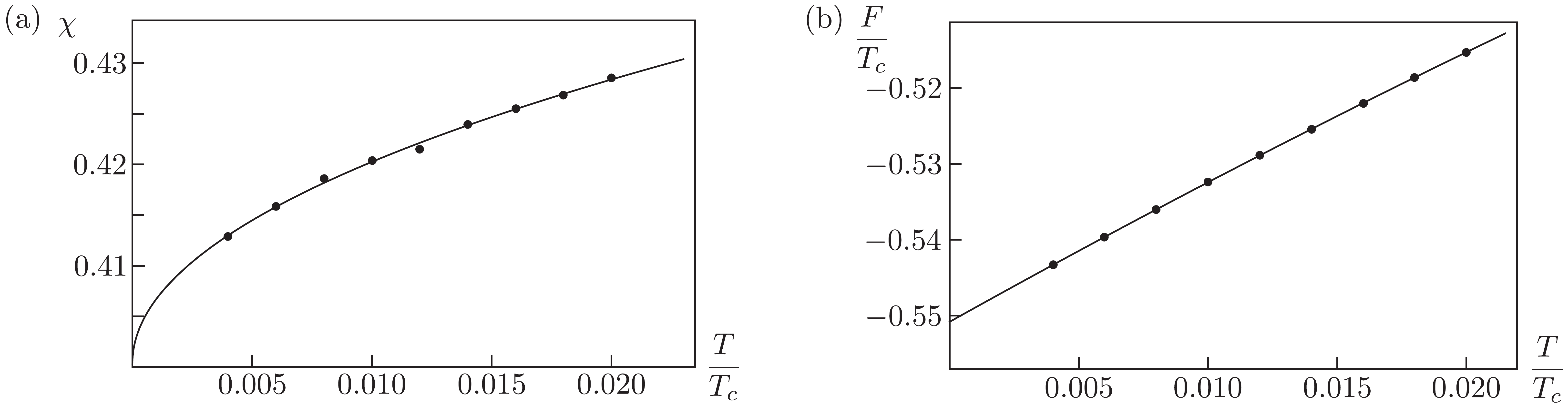}
  \end{center}
 \end{minipage}
\caption{(a)  Magnetic susceptibility in the low temperatures for $m^2=3.89$ and 
  the function (\ref{eq: chi low temp}) with $\chi_{0} = 0.401$ and $E_{0}=0.197$. (b)  Free energy in the low temperatures for $m^2=3.89$ and $\lambda=1$, and the function (\ref{eq: Fzero}) with $F_{0}=4.83 \times 10^{-4}$,  $G_{0} =1.67 \times 10^{-3}$, and $\gamma=5.44 \times 10^{-3}$.}  \label{fig5}
\end{figure*}

For the magnetization, we find that the temperature dependence in the low temperature region is approximately given by
\begin{equation}
M(T)\, \simeq\, M_{0} \left( 1 - D_{0}\, t ^{3/2}\right)\,. \label{eq: fitfn magzero}
\end{equation}
The fitting plot in the low temperature region is shown in Fig.\,\ref{fig4} for $m^{2} = 3.89$. The power law reduction of the magnetization at low temperatures, which goes as $\sim T^{3/2}$, is known as the Bloch $T^{3/2}$ law \cite{Ashcroft}. This dependence originates from the existence of spin 
wave excitations (or magnons) and shows that spin wave approximation is valid for the low-temperature ferromagnets. Our results (\ref{eq: fitfn magzero}) are consistent with this behavior, and the holographic dual model also reproduces the Bloch's law for magnetizations. This dependence indicates that our system is indeed dual to a ferromagnetic system, not an anti-ferromagnetic system. Fig. \ref{fig1} shows that the numerical results of $M(T)$ in the near-$T_c$ and the low-temperature regions are smoothly connected in our holographic calculation. The results regarding to the magnetization indicate that our holographic dual model in Eq. (\ref{eq: Holo action}) is capable of consistently describing both the mean field approximation near $T_c$ and spin wave approximation at low temperatures in a unified way. 

As for the magnetic susceptibility, the behavior in the low temperature region is well fitted with the following function:
\begin{equation}
\chi_{H}(T)\, \simeq\, \chi_{0}  +  E_{0}\, \sqrt{t}\,,   \label{eq: chi low temp}  
\end{equation}  
where $E_{0}$ is some constant. The numerical results for $m^2=3.89$ is summarized 
in Fig.\,\ref{fig5} $(a)$. The first constant term $\chi_0$ in Eq. (\ref{eq: chi low temp}) can be interpreted as the Pauli paramagnetic susceptibility from free electrons, and the fractional power law behavior of the second term implies the contribution from magnons. Even without any fermionic field on the action in Eq. (\ref{eq: Holo action}), the effect of fermions on the ferromagnetic side, which leads to the constant part of the susceptibility (i.e. conduction electron), can be possibly captured by our holographic model in the level of an effective theory. 

Finally, the temperature dependence of free energy at low temperatures is shown in Fig.\,\ref{fig5} (b) for $m^2 = 3.89$. The results for free energy $F(T)$ from the holographic dual model can be fitted with the following form,
\begin{equation}
F(T)\, \simeq\, -  F_{0}  +  G_{0}\, t  -  \gamma\, t^{2}\,, \label{eq: Fzero}
\end{equation}
where $F_{0}$, $G_{0}$, and $\gamma$ are positive constants. Here, $\gamma$ gives the coefficient linear in $T$ of the specific heat. This temperature dependence for the low-temperature region is consistent with the standard linear dependence of the specific heat for free electrons (or for the Fermi liquid). Interestingly, this result also indicates the existence of conduction electrons in our holographic dual ferromagnets. 

The low temperature behaviors in Eq. (\ref{eq: fitfn magzero}), (\ref{eq: chi low temp}), and (\ref{eq: Fzero}) indicate our holographic dual model (\ref{eq: Holo action}) actually corresponds to the ferromagnetic system with not only localized spins but also conduction electrons, such as models with the $s$-$d$ exchange interaction \cite{Ziman}.  Although the analysis in this Letter is performed based on the probe approximation, where the back reaction from the scalar field $\Phi$ is ignored, we expect the qualitative feature of the above results will be robust beyond this approximation. Indeed, in the limit of $\kappa^2/\lambda \ll 1$, we can safely neglect the contribution of $\Phi$ to the energy-momentum tensor in the Einstein equation and our analysis is justified even in the low temperature region, as discussed in Ref. \cite{Iqbal}.

\section{Summary and Discussion}
In summary, we have developed a holographic dictionary between a $(4+1)$-dimensional  gravitational system and a $(3+1)$-dimensional ferromagnetic system. Using the holographic dictionary, we numerically calculated relevant physical quantities, such as magnetization, magnetic susceptibility, and free energy. We verified that this holographic model produces these physical quantities consistent with the mean field theory near $T_c$. For low temperatures, we found the Bloch's $T^{3/2}$ law indicating the presence of the spin wave excitations, and moreover, the susceptibility and free energy exhibited signatures of conduction electrons. These results at the low temperatures indicate that our holographic model automatically captures the entire spectrum of ferromagnetic systems incorporating both spin and electronic degrees of freedom, and thus this model can develop into new effective descriptions of ferromagnetic systems useful for spintronics.

\section{Acknowledgement}
This work is supported by World Premier International Research Center Initiative (WPI), MEXT, Japan.
Y. N. and E. S. are supported by Grant-in Aid for Scientific Research on Innovative Areas "Nano Spin Conversion Science" (26103005). The work was supported by JSPS KAKENHI Grant No. 15K20877 for M. I. and No. 15K13531 for K. S.

\end{document}